\documentstyle[12pt]{article}

\begin{document}

\begin{center}
{\Large Derivative-Coupling Models and the Nuclear-Matter Equation of
State }\\
\vspace{1.5cm}
A. Delfino$^{\rm a}$, M. Chiapparini$^{\rm b,c}$, M. Malheiro$^{\rm a}$,
L. V. Belvedere$^{\rm a}$ and A. O. Gattone$^{\rm c}$\\
\vspace{0.2cm}
$^{\rm a}$ {\em Instituto de F\'{\i}sica - Universidade Federal
Fluminense}\\
{\em Outeiro de S\~ao Jo\~ao Batista s/n, \,24020-004 Centro,
Niter\'oi}\\
{\em Rio de Janeiro, Brazil}\\
\vspace{0.3cm}
$^{\rm b}$ {\em Departamento de F\'{\i}sica Nuclear e Altas
Energias}\\
{\em Centro Brasileiro de Pesquisas F\'{\i}sicas}\\
{\em Rua Xavier Sigaud 150, 22290-180 Rio de Janeiro RJ, Brazil} \\
\vspace{0.3cm}
$^{\rm c}$ {\em Departamento de F\'{\i}sica, TANDAR, Comisi\'on Nacional
de Energ\'{\i}a At\'omica,} \\
{\em Av. del Libertador 8250, 1429 Buenos Aires,
Argentina}

\end{center}
\vspace{1.0cm}

\begin{abstract}
The equation of state of saturated nuclear matter is derived using two
different derivative-coupling Lagrangians.  We show that both
descriptions are equivalent and can be obtained from the
$\sigma-\omega$ model through an appropriate rescaling of the coupling
constants.  We introduce generalized forms of this rescaling to study
the correlations amongst observables in infinite nuclear matter, in
particular, the compressibility and the effective nucleon
mass.
\end{abstract}

PACS number(s): 21.65.+f, 21.30.+y, 97.60.Jd\\
\vspace{1.0cm}

\section{Introduction}

The equation of state (EOS) of dense matter is known, experimentally,
at just one density value $\rho_0=0.15$/fm$^3$ (symmetric
nuclear matter) for a binding energy per particle $E/A(\rho_0)=-16$
MeV. The compression modulus is more loosely determined ranging from
as low as $\kappa=100$ MeV~\cite{brown-osnes} to the highest estimate
$\kappa=344$ MeV~\cite{co}. The last piece of experimental information
is the effective mass which, from the energy dependence of the
proton-nucleus optical potential, can be constrained to lie in the
range $M^*/M$=0.6---0.9.

Theoretically, the EOS is obtained using as starting point either
non-relativistic interactions of the Skyrme
type~\cite{jaminon-mahaux}, or relativistic mean-field models (RMF) of
baryons and mesons~\cite{walecka,brockman,fritz,haddad}.  As an
example of the non-relativistic calculations, the {\em
Skm*}-interaction~\cite{zimanyi}, used successfully to reproduce
ground-state nuclear properties, gives $M^*/M=0.78$ and a
compressibility $\kappa=217$ MeV. As for the RMF calculations, the
original linear $\sigma-\omega$ model~\cite{walecka} gives for the
same observables, a smaller effective mass $M^*/M=0.556$ and a larger
compressibility $\kappa=540$ MeV. The relativistic treatment of the
EOS becomes necessary at extreme conditions of density and temperature
such as those found in heavy-ion reactions at energies above 400
MeV/nucleon and type-II supernovas.  In what follows we will
concentrate on it.

The linear $\sigma-\omega$ (Walecka) model satisfactorily explains
many properties of nuclear matter and finite nuclei with two free
parameters.  The resulting compressibility at saturation density,
however, exceeds the experimental bound.  A way out of this difficulty
is to introduce non-linear scalar
self-couplings~\cite{sharma,glendenning}.  The resulting non-linear
$\sigma-\omega$ model reproduces well ground state nuclear properties
and is renormalizable (though it shows some instabilities at high
densities for low values of the compression modulus, $\kappa<200$
MeV).  The number of free parameters in this case is four.

One alternative approach which renders a satisfactory compression
modulus without increasing the number of parameters is that advanced
in Refs.~\cite{waldhauser,heide,zm}.  It employs the same degrees of
freedom and the same number of independent couplings present in the
Walecka model.  The difference is that it introduces a
non-renormalizable derivative coupling to the baryon field later
adjusted to reproduce the experimental conditions at saturation.  The
results for the compression modulus and effective mass compare well
with the Skyrme-type calculations.  The spin-orbit splitting in
finite-nuclei, however, turns out to be smaller than required by the
data.  We shall refer to this description as Model I.

A qualitative different model, though similar in spirit, obtains if
instead of modifying the covariant derivative only the baryonic
kinetic energy term is redefined.  As in Model I, a suitable rescaling
of the baryonic field leads to a Lagrangian describing baryons with
effective mass $M^*$.  Such a model was originally suggested as a
possibility by Zimanyi and Moskowski in Ref.~\cite{zm} (appendix) and
implemented for a particular case by Delfino {\em et
al}.~\cite{delfino}.  The experimental nuclear matter saturation
density and binding energy per nucleon are used to fit the two free
parameters; the compressibility comes out rather small but within the
bounds (156 MeV) and the spin-orbit splitting doubles the value
obtained in Model I (though still short of the data).  To this model
we shall refer hereafter as Model II.

The aim of this paper is to provide a unified discussion of models I
and II.  We will show that both are to be understood as generalized
Walecka models where the meson couplings ($g_\sigma$ and $g_\omega$)
become {\em effective} coupling constants.  In Model I only the scalar
coupling is affected; in Model II both scalar and vector couplings are
modified.  The discussion will lead us naturally to consider a
generalization of the models above effected by the introduction of a
third parameter $\alpha$ which rescales the effective coupling
constants.  We will explore the results for $M^*$, $\kappa$, and the
mean-field vector (V) and scalar (S) potentials, within both models
for different values of the parameter $\alpha$.

Our basic motivation is to sum up in a phenomenological meson-baryon
model, with as few parameters as possible, the ability to reproduce
the four nuclear-matter observables and to give a good description of
ground-state properties of finite nuclei without the presence of
instabilities at high densities.  It is not a substitute to the more
fundamental relativistic Brueckner-Hartree-Fock
(RBHF) method~\cite{DBHF,li,terhaar}  but a simplifying alternative to
explore the nuclear matter equation of state.

It is important to stress that the same idea of density-dependent
coupling constants underlies two recent approaches.  The first one,
known as relativistic density-dependent Hartree-Fock~\cite{shi},
describes finite nuclei and nuclear matter saturation properties using
coupling constants that are fitted, at each density value, to the RBHF
self-energy terms.  The good agreement obtained for the ground state
properties of spherical nuclei lends support to a description
involving these density-dependent coupling constants.  In the second
approach~\cite{miyazaki}, a similar density dependence of the
parameters is achieved through projection of the meson-nucleon
vertices into positive-energy space weighted with a parameter adjusted
to reproduce saturation properties.  Here also, the generalization of
the model through the inclusion of adjustable parameters at the
vertices permits to study the density dependence in an analytical
manner.  Both approaches have natural partners in the relativistic
density-dependent Hartree-Fock of Li and Zhuo~\cite{li1} and in the
Gmuca's model~\cite{gmuca}.  Our treatment of models I and II belongs
to this category of descriptions involving density-dependent coupling
constants.

The outline of the paper is as follows: in the next section we present
the two models, establish their form and relationship and give
expressions for the generalized EOS and compressibility. In section 3
we present our results and discuss the correlation between the
calculated nuclear matter observables. Finally, in the last section
we present our conclusions.

\section{The Models}

\indent We begin by introducing the non-linear Lagrangian density
of Model I,
\begin{equation}
{\cal L}_{NL}=\bar \psi \left\{ \slash\!\!\!\!D(g_{\omega})-
m^{\ast}(\sigma)M \right\} \psi
-\frac{1}{4} F^{\mu \nu} F_{\mu \nu}+ \frac{1}{2} m^2_{\omega}
\omega_{\mu}
\omega^{\mu}+\frac{1}{2}\left(\partial_{\mu }\sigma
\partial^{\mu}\sigma -m^2_{\sigma}\sigma^2 \right)\,,   \label{3}
\end{equation}
where $\slash\!\!\!\!D(g_{\omega}) \doteq \gamma^{\mu}
D_{\mu}(g_\omega) = \gamma^{\mu}(i\partial_{\mu }- g_{\omega}
\omega_{\mu })$ is the usual covariant derivative and the degrees of
freedom are the baryon field $\psi$, the scalar-meson field $\sigma$,
and the vector-meson field $\omega^{\mu}$.  The real function
$m^{\ast}(\sigma)$ is unknown except for the fact that it must go to
unity as the density goes to zero and has to vanish at large densities
where the effective mass approaches zero asymptotically.  Indeed, the
Dirac equation obtained from the Lagrangian density (\ref{3}) gives
\begin{equation}
m^{\ast}({\sigma})=\frac{\phantom{1}M^*}{M}\,, \label{4}
\end{equation}
where $ M $ and $ M^{\ast} $ are the bare and
effective baryonic mass, respectively.

We show next that performing a transformation on the spinor field
${\cal L}_{NL}$ can be derived from a Lagrangian density with
derivative scalar coupling (DSC). The proposed Lagrangian density
\cite{zm} which generates ${\cal L}_{NL}$ is given by
\begin{equation}
{\cal L}=\bar \psi \left\{ [m^{\ast}(\sigma)]^{-
1}\slash\!\!\!\!D(g_{\omega})-M \right\} \psi -\frac{1}{4} F^{\mu
\nu} F_{\mu \nu} + \frac{1}{2} m^2_{\omega}\omega_{\mu}
\omega^{\mu}+\frac{1}{2}\left(\partial_{\mu }\sigma
\partial^{\mu}\sigma - m^2_{\sigma} \sigma^2 \right)
\,. \label{5}
\end{equation}
Introducing the rescaled baryonic field  $ \psi \rightarrow
[m^{\ast}(\sigma)]^{\frac{1}{2}}\psi $, we obtain from Eq.(\ref{5})
the rescaled Lagrangian density
\begin{equation}
{\cal L}_{R}={\cal L}_{NL}+{\Im}\,, \label{6}
\end{equation}
where $ \Im $ is an imaginary contribution given by
\begin{equation}
{\Im}=\frac{i}{2}(\bar \psi \gamma_{\mu}\psi)
\partial^{\mu}\ln (m^{\ast}(\sigma))\,. \label{7}
\end{equation}
This term does not carry any physical content and can be transformed
away by substituting the baryonic kinetic energy term in Eq.(\ref{3})
for the symmetric derivative $ \frac{i}{2} \{ \bar \psi \gamma_{\mu}
\partial^{\mu } \psi - (\partial^{\mu } \bar \psi )\gamma _{\mu} \psi
\} $. The imaginary contribution $ \Im $ cancels after the field
scaling.  Once performed the field rescaling is equivalent to the
replacement,
\begin{equation}
\left\{[m^{\ast}(\sigma)]^{- 1}\slash\!\!\!\!D(g_{\omega})-
M \right\} \rightarrow
\left\{ \slash\!\!\!\!D(g_{\omega})-
m^{\ast}(\sigma)M\right\}\,.\label{8}
\end{equation}
The Lagrangian densities given by Eqs.(\ref{3}) and (\ref{5}) are,
therefore, completely equivalent; they describe the same physics
whether we deal with infinite nuclear matter or finite nuclei.  The
equations of motion obtained from Eqs.(\ref{3}) and (\ref{5}) give
rise to the same hadronic dynamics.  The modified kinetic energy of
Eq.(\ref{5}) is not arbitrary; it describes the motion of a baryon of
mass $ M^{\ast} $ instead of the bare mass $ M $. Equation~(\ref{3})
just carries this information to the scalar-baryon coupling fields.

The Walecka model can be obtained as a particular case of either
Lagrangian density by making the choice $
m^{\ast}(\sigma)=(1-g_{\sigma}{\sigma}/M) $. More generally, recent
nonlinear models like those presented in
Refs.~\cite{zm,greiner,feldmeier,koepf} (which are variations of the
DSC model) may be interpreted as modified Walecka models where,
\begin{equation}
{\cal L}_{NL} \equiv {\cal L}_{Walecka}(g_{\sigma}\rightarrow g_{\sigma}
^{\ast})\,,\label{9}
\end{equation}
and $ g_{\sigma}^{\ast} $ (hereafter we will interpret $\ast$ as
refering to effective coupling constants in the medium) is now a
function of $ \sigma $ related to $ m^{\ast}(\sigma) $ by
\begin{equation}
m^{\ast}(\sigma)=1 - g_{\sigma}^{\ast}\sigma/M\, . \label{10}
\end{equation}
By itself, this establishes a class of models since
$m^{\ast}(\sigma)$ is still a generic real function.  In the usual
Zimanyi-Moszkowski (ZM) model~\cite{zm}, for example,
\begin{equation}
m^{\ast}_{ZM}(\sigma)=(1 + g_{\sigma}{\sigma}/M)^{-1}\,,\label{11}
\end{equation}
and $ g_{\sigma}^{\ast}=g_{\sigma}m^{\ast}_{ZM}(\sigma)$.  Likewise,
the identification of $ g_{\sigma}^{\ast} $ and $ m^{\ast}(\sigma) $
for the other models~\cite{greiner,feldmeier,koepf} can be easily
done.

We introduce next a modified version of Model I that
we refer to as Model II. We keep the generalized
factor $m^{\ast}(\sigma)$
but, following the suggestion given in Ref.\cite{zm},
we restrict the $ m^{\ast}(\sigma) $ dependence in ${\cal L}$
to the fermionic kinetic energy term. To this end we define a
modified covariant derivative
$\,\slash\!\!\!\!D_{m^\ast}(g_{\omega})\equiv\left
([m^{\ast}(\sigma)]^{- 1} i\slash\!\!\!\partial -
g_{\omega}\slash\!\!\!\omega \right)$ and write the Lagrangian in the
form
\begin{equation}
{\cal L}=\bar \psi \left\{\slash\!\!\!\!D_{m^\ast}(g_{\omega})-
M\right\}\psi-\frac{1}{4} F^{\mu \nu} F_{\mu
\nu}+\frac{1}{2}m^2_{\omega}\omega_{\mu}
\omega^{\mu} +\frac{1}{2}\left(\partial_{\mu }\sigma
\partial^{\mu}\sigma -m^2_{\sigma}\sigma^2 \right) \,. \label{12}
\end{equation}
A rescaling similar to the previous one, $ \psi \rightarrow
[m^{\ast}(\sigma)]^{\frac{1}{2}}\psi $, transforms
\begin{equation}
\left\{\slash\!\!\!\!D_{m^\ast}(g_{\omega})-
M\right\}\rightarrow \left\{\slash\!\!\!\!D(g^{\ast}_{\omega})-
m^{\ast}(\sigma)M\right\}\,, \label{13}
\end{equation}
thus generating a new class of nonlinear models.
The connection with the Walecka model is an extension of the previous
one and proceeds through the change,
\begin{equation}
{\cal L}_{NL} \equiv {\cal L}_{Walecka} ( g_{\sigma}\rightarrow
g_{\sigma}^{\ast};\: g_{\omega}\rightarrow
g_{\omega}^{\ast})\,.\label{14}
\end{equation}
Here $ g_{\sigma}^{\ast} $ is connected with $m^{\ast}(\sigma)$
by Eq.(\ref{10}) and
\begin{equation}
\frac{g_{\omega}^{\ast}}{g_{\omega}} = m^{\ast}(\sigma)\,. \label{15}
\end{equation}
Equations~(\ref{9}) and (\ref{14}) help to understand the
different types of nonlinear couplings showing that they are, in fact,
effective Walecka models. An important point to emphasize is that
given the Lagrangian of Eq.(\ref{12}), the
scaling of the effective vector coupling constant in the medium is
fixed by (\ref{15}).

We now explore the possibility of extending the same scaling to the
effective scalar coupling constant in the medium.  Thus, we look for a
function $ m^{\ast}(\sigma) $ such that the following constraint is
satisfied:
\begin{equation}
\frac{ g_{\sigma}^{\ast}}{ g_{\sigma}} =
 \frac{ g_{\omega}^{\ast}}{ g_{\omega}}=m^{\ast}\,\,.\label{16}
\end{equation}
It turns out that the only function that fulfills this
requirement is $m^{\ast}(\sigma) = m^{\ast}_{ZM}(\sigma)$.
We call the Model II Lagrangian with $m^{\ast}(\sigma) =
m^{\ast}_{ZM}(\sigma)$ the
modified ZM model (MZM) and is the first hadronic model which
exhibits this property. It couples the scalar $ \sigma $
to the vector $ \omega $ field and gives results for the nuclear
matter observables substantially different from
the usual ZM model. A comparison follows.

We took for the usual ZM model, as given by Eq.(\ref{9}), the coupling
constants used in Ref.\cite{zm}.  For the MZM model, Eq.(\ref{14}), we
fit the parameters to the saturation density and the experimental
binding energy per nucleon to obtain, $ C_{\sigma}^2 \equiv
g_{\sigma}^2M^2/m_{\sigma}^2 = 443.3 $ and $ C_{\omega}^2 \equiv
g_{\omega}^2M^2/m_{\omega}^2 =305.5$~\cite{delfino}.  In the table
below we show, for both models, the numbers obtained for the
compressibility $\kappa$, the scalar potential $S$, the vector
potential $V$ (in MeV) and the baryonic effective mass $m^{\ast}$:

\vspace{0.5cm}

\begin{center}
\begin{tabular}{|c|c|c|c|c|} \hline\hline
 Model   &  $m^*$  &  $\kappa$  &  {\em S}  &  {\em V} \\ \hline
 ZM      &   0.85  &    225    &   -141    &    82    \\
 MZM     &   0.72  &    156    &   -267    &    204   \\ \hline\hline
\end{tabular}
\end{center}

\vspace{0.5cm}

Compared to ZM, the MZM model gives, simultaneously, a smaller
$m^{\ast}$ and a smaller $\kappa$.  This is unlike what is observed
when comparing ZM with the Walecka model.  This difference in
predictions is explained by the nonlinear scalar-vector coupling
contained in the MZM Lagrangian.  Furthermore, it is known that the ZM
model gives poor results for the spin-orbit splitting (which is
strongly dependent on the quantity $V - S$) when used in finite nuclei
calculations~\cite{koepf}. In the MZM model, this quantity more than
doubles that of ZM, suggesting that MZM may improve upon the former in
this particular direction.

It is also interesting to remark that at low energies the slope of the
real optical potential (given by $ 1 - m^\ast $) provides information
regarding the expected value of $m^\ast$.  The experimental value, in
the limit of infinite mass number and zero radius, gives $m^*$ around
0.6~\cite{feldmeier,perey}.  This result does not lend support to the
ZM model, but tend to favour the Walecka and MZM models instead
\cite{delfino}.  Finally, $\kappa$ is smaller in the MZM model than
the ``empirical'' prediction $ \kappa = 210 \pm 30 $ MeV. Despite
this, the advantage of MZM here is that, unlike the nonlinear $\sigma
- \omega$ model~\cite{boguta}, it does not present anomalies in the
EOS for any value of $\kappa$.

We proceed, now, to generalize the ZM and the MZM models.  We recall
that both originate from Eq.(\ref{10}) for the particular choice
$m^{\ast}(\sigma)= m^{\ast}_{ZM}(\sigma)\, $. As a consequence of this
choice the scalar effective coupling constant scales as $ (
g_{\sigma}^{\ast}/ g_{\sigma}) = m^{\ast}$.  However, different
choices of $m^{\ast}(\sigma)$ can be made meeting the requirement of
Eq.(\ref{10}).

\subsection{Model I}

\indent
From Eq.(\ref{9}) we generate a family of models by choosing a scaling
given by \begin{eqnarray} {\rm Model\;I:}
&g_{\sigma}^{\ast}/g_{\sigma}= m^{{\ast}^{\alpha}}
&g_{\omega}^{\ast}/g_{\omega}= 1 \label{1} \end{eqnarray} The equation
of state for such a family of models will be shown in the next section
in a general expression along with that of Model II.  For Model I we
consider two particular cases:  the Walecka model ($\alpha = 0$) and
ZM model ($\alpha = 1 $).

\subsection{Model II}

\indent Defined by Eq.(\ref{14}) and the scaling
\begin{eqnarray}
{\rm Model\;II:}     &g_{\sigma}^{\ast}/g_{\sigma}=
m^{{\ast}^{\alpha}} &g_{\omega}^{\ast}/g_{\omega}= m^* \label{2}
\end{eqnarray}
(the MZM model is obtained for $\alpha$ = 1).  For a given value
of $\alpha$, the function $m^{\ast}(\sigma)$ is defined by
Eq.(\ref{2}) together with (\ref{10}) as in the Model I case.

\subsection{The Equations of State for the Models}

\indent
When the meson fields in the Lagrangian of both models are substituted
by their mean values we arrive at the mean field approximation (MFA).
In Model I the scalar-meson field $\sigma$ is a function of the scalar
density ($\rho_s$) exclusively.  In Model II, $\sigma$ depends on
$\rho_s$ but also on the baryonic density $\rho_{b}$.  For
rotationally and translationally invariant symmetric nuclear matter,
the MFA equation for the scalar fields reads
\begin{equation}
\sigma=\frac{{g_\sigma}}{m_\sigma^2M}\frac{{m^*}^{\alpha+1}}
{(1-\alpha)m^*+\alpha}\left[M\rho_s +
\beta\left(\frac{g_\omega}{m_\omega}\right)^2
m^* \rho_b^2 \right]\,\,, \label{18}
\end{equation}
with $\beta=0$ or 1 for Model I or Model II, respectively.
Equation~(\ref{18}) shows clearly how Model II and its generalization
mix the scalar and vector fields.  The scalar and baryonic densities
are related through,
\begin{equation}
\frac{\rho_s}{\rho_b}=-\left(\frac{C_{\omega}^2}{C_{\sigma}^2}\right)
\left(\frac{(1-\alpha)m^\ast +
\alpha}{{m^*}^{2\alpha+1-2\beta}}\right)
\left(\frac{S}{V}\right)
-\beta\left(\frac{V}{Mm^\ast}\right)\,\,, \label{19}
\end{equation}
where
\begin{equation}
S =-g_{\sigma}^{\ast}\sigma=-M(1 - m^*)\,, \label{20}
\end{equation}
and
\begin{equation}
V=(C_{\omega}^{2}/M^2){m^*}^{2\beta}/{\rho_b}\,. \label{21}
\end{equation}
This density ratio weighs the content of each model.  Model II has an
additional term making its contribution to the $\sigma$-field, and
consequently to the effective mass, larger than that of Model I.

The expressions for the energy density and pressure at a given
temperature $ T $ can be found, as usual, by the MFA average of the
energy-momentum tensor. They read,
\begin{equation}
{\cal E} = \frac{C_{\omega}^{2}}{2M^{2}}m^{\ast^{2\beta}}\rho_{b}^2+
\frac{M^{4}}{2C_{\sigma}^{2}}\left(\frac{1-m^{\ast}}{m^{\ast{\alpha}}}
\right)^2+ \frac{\gamma}{(2\pi)^{3}}\int
d^{3}k\,E^{\ast}(k)(n_{k}+\bar n_{k})
\,, \label{22}
\end{equation}
and
\begin{equation}
p=\frac{C_{\omega}^{2}}{2M^{2}}m^{\ast^{2\beta}}\rho_{b}^2-
\frac{M^{4}}{2C_{\sigma}^{2}}\left(\frac{1-m^{\ast}}
{m^{\ast{\alpha}}}\right)^2
+\frac{1}{3}\frac{\gamma}{(2\pi)^{3}}\int
d^{3}k\,\frac{k^{2}}{E^{\ast}(k)}
(n_{k}+\bar n_{k})\,,  \label{23}
\end{equation}
where
\begin{equation}
\rho_{b} = \frac{\gamma}{(2\pi)^{3}}\int d^{3}k\,(n_{k}-\bar
n_{k})\,.\label{24}
\end{equation}
Here $ \gamma$ is the degeneracy factor ($\gamma = 4 $ for nuclear
matter; $ \gamma = 2 $ for neutron matter), $\bar n_{k}$ and
$n_{k}$ stand for the Fermi-Dirac distribution for antibaryons and
baryons with exponent $ (E^{\ast} \pm \nu)/T $, respectively. The
energy $E^{\ast}(k)$ is given by
\begin{equation}
E^{\ast}(k) = ( k^2 + M^{\ast 2} )^{\frac{1}{2}}\,, \label{25}
\end{equation}
while the effective chemical potential which preserves the number of
baryons and antibaryons in the ensemble is defined by \,$  \nu=\mu -
V$  , with $\mu$ the thermodynamical chemical potential.
According the Hugenholtz-van Hove theorem \cite{hugenholtz}, the
Fermi energy must be equal to the energy per baryon at saturation
density. Therefore, the following relation has to be satisfied,
\begin{equation}
\frac{ {\cal E}}{\rho_{o}} = V +  E^{\ast}(\rho_{o})\,.\label{26}
\end{equation}

We end the section by presenting an analytical general expression
for the compressibility valid in both models,
\begin{equation}
K=9V+ 3\frac{k_{F}^{2}}{(k_{F}^2+(M+S)^2)^{1/2}}
+9\left(\rho\frac{\partial S}{\partial\rho}
  \left(\frac{(M+S)}{(k_{F}^2+(M+S)^2)^{1/2}}+2\beta
  \frac{V}{(M+S)}\right)
      \right)\,,
      \label{27}
\end{equation}
where $k_F$ is the Fermi momentum and all quantities are calculated at
nuclear matter saturation density $\rho_o$.

\section{Results and Discussions}

\indent
We have conducted calculations with these generalized models at $T=0$,
for different values of $\alpha$, requiring $ E_b=-15.75 $ MeV at $
\rho_\circ = 0.15\,$fm$^{-3}$ . We begin this section by discussing
Model I. The question we ask ourselves is whether, by varying
$\alpha$, we can obtain a simultaneous fit for $\kappa$, $M^*$ and $V
- S$.  The answer to this question is negative.  This can be better
appreciated by looking at Fig.~1 where these quantities are plotted as
a function of $\alpha$.  No region in the plot gives a simultaneous
``reasonable" agreement for the discussed quantities.  By
``reasonable" we mean $m^*$ around 0.6 (which, we shall see later,
fixes the values of $V-S$\ at approximately 680 MeV) and a
compressibility $\kappa = 210\pm 30$ MeV which is the value of derived
from the energy of the breathing mode of doubly magic
nuclei~\cite{blaizot}.  This conclusion agrees with that of Greiner
and Reinhard~\cite{greiner} using a generalized ``ansatz" applied
to the same model.  A word about the nuclear compressibility; using
Fermi-liquid Landau theory, Brown~\cite{brown} has argued strongly in
favour of a lower value of $\kappa$.  Since the subject is still under
debate we use the value of Ref.\cite{blaizot} as an approximate upper
limit.

The results for Model II are shown in Fig.~2.  The compressibility
shows a weak dependence on $\alpha$ in the region of $\alpha\,>\,1$,
and reaches a minimum close to $\alpha=1$ which corresponds to the MZM
case.  Particulary interesting are the results for the region
$\,\alpha < 1$ where $m^{\ast}$ and the difference $ V - S $ show an
improvement over the results of the MZM model given in the previous
section.  For example, for $\alpha=0.9$ Model II gives
$M^{\ast}=558.3$ MeV, $V - S=687.6$ MeV and a compressibility
$\kappa=166.2$ MeV. It is also interesting to note that for
$\alpha=0.88$ one gets $M^{\ast}=507.0$ MeV, and $ V - S=785.1$ MeV
which are very close to the values obtained in the usual linear
Walecka model but, this time, with a softer compressibility,
$\kappa=181.3$ MeV.

Figure~2 shows also that for $\,\alpha <0.79 $ nuclear matter does not
saturate.  This includes the $\alpha{=}0$ case, which corresponds to
the Walecka model if $m^{\ast}(\sigma)$ is as in Eq.(\ref{10}).  In
addition, Figs.~1 and 2 indicate that $M^{\ast}$ increases with
$\alpha$ thus decreasing the vector potential due to (\ref{26}).  We
have allowed for higher values of $\alpha$ in the calculations to
reach the curious situation where $V$ vanishes, and model I and II
become degenerate.  Nuclear matter saturation is achieved with just
the scalar field.  This situation occurs for $\alpha \approx 12.8$,
which is the maximum value that this parameter can have (beyond it $V$
becomes negative), and for $C_{\sigma}^{2}=315.36$.  The results for
this special case are:  $M^{\ast}=885.8$ MeV, $ V - S=-S=52.4$ MeV and
$K=65.8$ MeV.

With some insight into the two models, we discuss here if in any
$M^*$ could give model-independent properties of nuclear matter.
As we argued before the effective mass is a manifestation of the
dynamical relativistic content of any particular model.  On the other
hand, if two models have the same $M^*$, they have the same density
ratio as given by Eq.(\ref{19}).  This quantity is shown in Fig.~3 as
a function of $\alpha$.  We notice that Model II, unlike Model I, can
acquire a strong relativistic character.  Going back to Figs.~1 and 2
we notice that the same $M^*$ can be obtained in both models for
different values of $\alpha$.  Consequently, they will give the same
relativistic ratio.  If the situation is such that two different
models give the same $M^*$, the question is whether other observables
can be determined likewise.  For instance, for a given $M^*$, the
potentials $S$ and $V$ are fixed by Eqs.(\ref{20}) and (\ref{26}),
respectively.  Therefore, the quantity $V-S$ and $M^*$ are directly
correlated and carry the same physical information for each particular
model.  As for the correlation between $\kappa$ and $M^*$, the
situation is different having a more model-dependent character.

In order to learn about the compressibility $\kappa$, we have
calculated its individual contributions according to Eq.(\ref{27}).
To the first three terms of (\ref{27}) we will refer as $\kappa_{1}$,
$\kappa_{2}$ and $\kappa_{3}$, respectively.  An additional fourth
term ($\kappa_{4}$) is present for Model II ($\beta=1$).  These
quantities are plotted in Figs.~4 and 5. $\kappa_{1}$ and $\kappa_{2}$
are completely determined by $M^*$ since they depend directly on $S$
and $V$; $\kappa_{3}$ and $\kappa_{4}$, however, show an extra
dependence on the slope of $M^{\ast}(\rho)$ which is negative and
carries information about the scalar field.  For Model I, Fig.~1 shows
that as $M^*$ increases the compressibility $\kappa$ decreases.  This
explains why, in the Walecka model where $M^*$ is small, $\kappa$ is
so high.  In the usual ZM model the opposite is true.  In Model II the
changes in $\kappa_3$ and $\kappa_4$ are such that despite the
increase in $M^*$ the total compressibility remains almost constant.

Finally, in Fig.6 we show the effective mass $m^\ast$ ---solution of
Eq.(\ref{10})--- for different values of $\alpha$.

\section{Conclusions}

\indent
In summary, we have shown the equivalence between two descriptions of
derivative scalar coupling models which we now interpret as effective
Walecka models.  In them the effective coupling constants depend on
the density and are completely determined by the effective nucleon
mass $ m^{\ast} $. The first model has only one effective coupling
$g^{\ast }_\sigma$ related to $m^{\ast }$ through $ m^{\ast } = 1 -
g^{\ast}_\sigma \sigma / M $. The second includes also an effective
vector coupling $ g^{\ast}_{\omega} $ which is always given by $
g^{\ast}_{\omega} = m^{\ast} g_{\omega} $. We have also shown that for
a particular choice of $ m^{\ast} = m^{\ast}_{ZM}(\sigma) $ both
effective meson coupling constants scale with $ m^{\ast} $. We have
generalized this scaling by introducing an extra parameter $\alpha$
and concluded that, in a theory with three free parameters, Model I
can not provide good results for the level splitting in finite nuclei
and at the same time adjust the experimental $\kappa$ and $m^*$.  We
showed, however, that it is possible to choose the parameter $\alpha$
so as to give values of $m^*$\,, $\kappa$\, and $ V- S $ similar to
those obtained from the nonlinear $\sigma- \omega$ model.  We conclude
that changing only the scalar coupling (Model I) definitely improves
$\kappa$ but with no hope of reproducing the spin-orbit splitting for
finite nuclei, thus supporting the conclusions of Ref.\cite{greiner}.
This is a direct consequence of the large values $M^{\ast}$ that Model
I gives when $\kappa$ goes in the right direction as ${\alpha}$
changes.  Model II, instead, through the inclusion of a mixed coupling
between the scalar and the vector fields may be the way to expect
improvement in the calculation of the observables.

\vspace{1cm}

{\bf Acknowledgements}

The authors would like to express their thanks to the Conselho
Nacional de Desenvolvimento Cient\'{\i}fico e Tecnol\'ogico (CNPq),
Brazil, for its financial support.  M.C. is a fellow of the
the Centro Latinoamericano de F\'{\i}sica (CLAF).

\newpage

\begin{figure}
\caption{
The nucleon effective mass ($M^*$), the difference between
the vector and scalar potentials ($V-S$) and the compressibility
$\kappa$ as function of $\alpha$ for Model I.}
\end{figure}
\begin{figure}
\caption{
The nucleon effective mass ($M^*$), the difference between
the vector and scalar potentials ($V-S$) and the compressibility
$\kappa$ as function of $\alpha$ for Model II.}
\end{figure}
\begin{figure}
\caption{
The relativistic ratio between scalar and baryonic densities
($\rho_s /\rho_b $) as a function of $\alpha$ for both models.}
\end{figure}
\begin{figure}
\caption{
The components $\kappa_1$, $\kappa_2$ and $\kappa_3$ of the
compressibility $\kappa$ as a fuction of $\alpha$ for Model I.}
\end{figure}
\begin{figure}
\caption{
The components $\kappa_1$, $\kappa_2$, $\kappa_3$ and $\kappa_4$ of the
compressibility $\kappa$ as a fuction of $\alpha$ for Model II.}
\end{figure}
\begin{figure}
\caption{
The effective nucleon mass as a function of the scalar
sigma field ($ u=g_{\sigma}\sigma/M$) for different values of
$\alpha$.}
\end{figure}
\end{document}